\documentclass[aip,reprint]{revtex4-1}

\usepackage{graphicx}

\begin{document}

\title{Transient phases during fast crystallization of organic thin films from solution} 



\author{Jing Wan}
\author{Yang Li}
\author{Jeffrey G. Ulbrandt}
\affiliation{Department of Physics and Materials Science Program, University of Vermont, Burlington VT 05405}
\author{Detlef-M. Smilgies}
\affiliation{Cornell High Energy Synchrotron Source, Cornell University, Ithaca NY 14853}
\author{Jonathan Hollin}
\author{Adam C. Whalley}
\affiliation{Department of Chemistry, University of Vermont, Burlington VT 05405}
\author{Randall L. Headrick}
\email[]{rheadrick@uvm.edu}
\affiliation{Department of Physics and Materials Science Program, University of Vermont, Burlington VT 05405}


\date{\today}

\begin{abstract}
We report  an in-situ microbeam grazing incidence X-ray scattering  study of 2,7-dioctyl[1]benzothieno[3,2-b][1]benzothiophene (C$_8$-BTBT) organic semiconductor thin film deposition by hollow pen writing. Multiple transient phases are observed during the crystallization for substrate temperatures up to $\approx$93$^\circ$C.    The layered smectic liquid-crystalline phase of C$_8$-BTBT initially forms and preceedes inter-layer ordering, followed by a transient crystalline phase for temperature $>$60$^\circ$C, and ultimately the stable phase.   Based on these results, we demonstrate a method to produce extremely large grain size and high carrier mobility during high-speed processing. For high writing speed (25 mm/s) mobility up to 3.0 cm$^2$/V-s has been observed.
\end{abstract}


\maketitle 

\section{Introduction}

Solution-processed organic semiconductor thin films have attracted great interest due to their potential applications in low-cost and flexible organic electronic devices\cite{Sirringhaus2000}. An important challenge lies in the manipulation of morphology and crystalline ordering of molecules, which critically influences the electronic properties of thin films.\cite{Lim:2009aa,kumar2014}   However, due to the weak Van Der Waals forces between organic semi-conducting molecules, the molecular packing depends  sensitively on the processing methods and conditions.\cite{Diao2014aa}  Therefore, understanding the crystallization mechanisms, which turn out to be both subtle and varied, is important to give insight into controlling the deposition processes.

We address the question of whether high-speed pen writing can be optimally used for the fabrication of electronic devices such as organic field effect transistors (OFETs). At high enough speeds, films are deposited in a liquid state that subsequently transforms to a solid due to evaporation of the solvent; this is known as  the Landau-Levich-Derjaguin (LLD) regime.\cite{landau1942,derjaguin1943}   Although LLD extends to practically unlimited writing speeds,  LLD normally leads to a high nucleation density resulting in an isotropic small-grain structure.\cite{wo2012} This results in grain boundaries and other defects, which introduce trap states that compromise charge carrier mobility.   Hence, oriented single-crystalline films are desirable for achieving low defect density, but this normally requires slow writing in the ``convective" deposition regime.\cite{headrick2008,cour2013origin}  Thus, there is seemingly no optimal strategy for high-speed processing. Here, we report the discovery of mechanisms that lead to large grain size in the LLD  regime, effectively solving this long-standing problem.

%
%

\begin{figure}
\includegraphics[width=3.5 in]{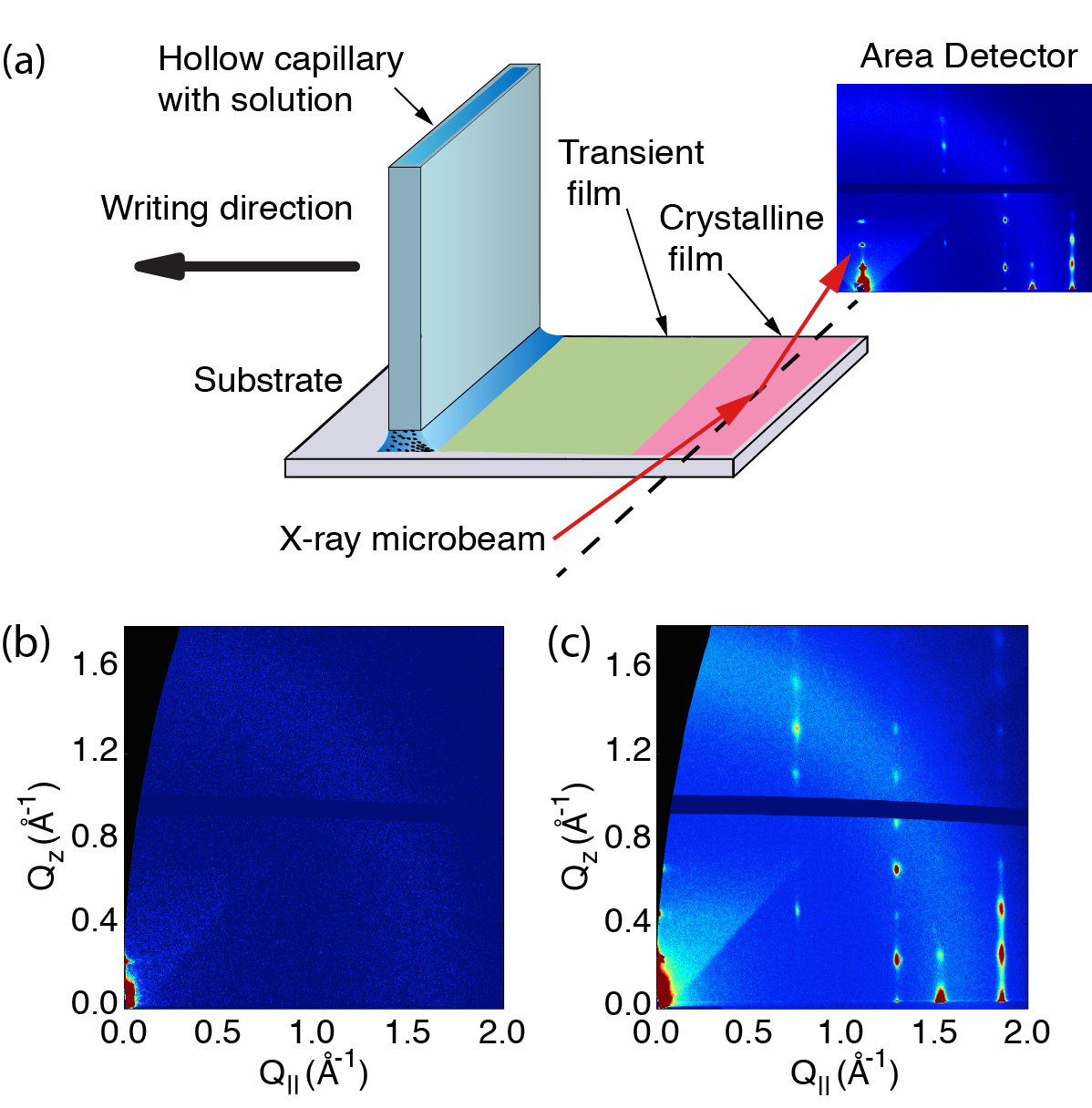}
\caption{\label{Expt_Layout} (a) Schematic of the $\mu$GIWAXS real-time pen-writing system.  (b) Scattering pattern of the transient liquid-crystalline state during writing at  85$^\circ$C captured $<$1 s after the pen has passed through the scattering region. Only the  specular (001) at $Q_z \approx 0.22$ {\AA}$^{-1}$ is present. (c) $\mu$GIWAXS image of the final crystalline state (Cr2), showing complete development of the in-plane structure.}
\end{figure}

\begin{figure}
\includegraphics[width=3.0 in]{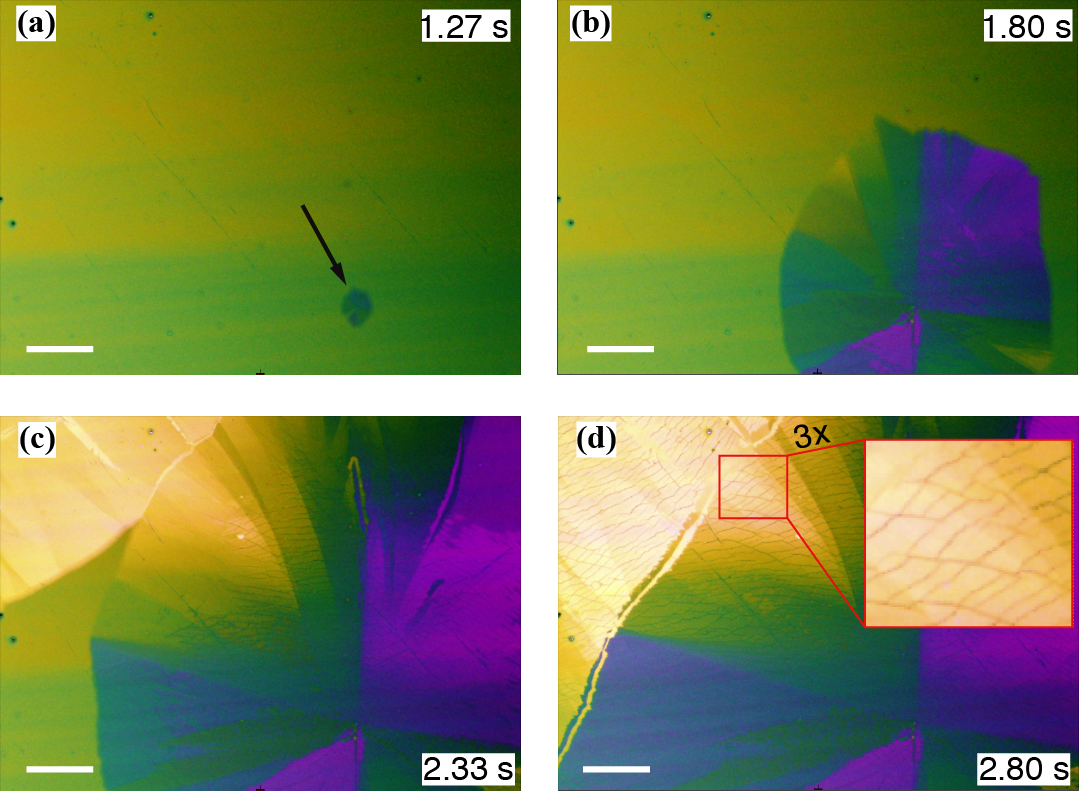}
\caption{\label{optical} Frames from a polarized optical  microscopy video during deposition at 90$^\circ$C with 25 mm/s writing speed. The translation stage stops at $\Delta t = 0.47$ s after the region of the film shown is written. A featureless view remain until $\Delta t$ = 1.27 s as shown in (a): a nuclei at the right bottom appear as indicated by the arrow. (b) The crystallization propagates from the nucleus center. No cracks are visible on the film in this image. (c) Cracks forming from the nucleus center spread quickly. (d) The cracks are fully formed and some minor color change has occurred near the grain boundary. The relative time $\Delta t$ is labeled on the right bottom of the frames and the scale bar of 200 $\mu$m is labeled on the left bottom. See Supplementary Movie 1 for the complete sequence of images.\cite{Supplementary}}
 \end{figure}

\begin{figure*}
\includegraphics[width=5.25 in]{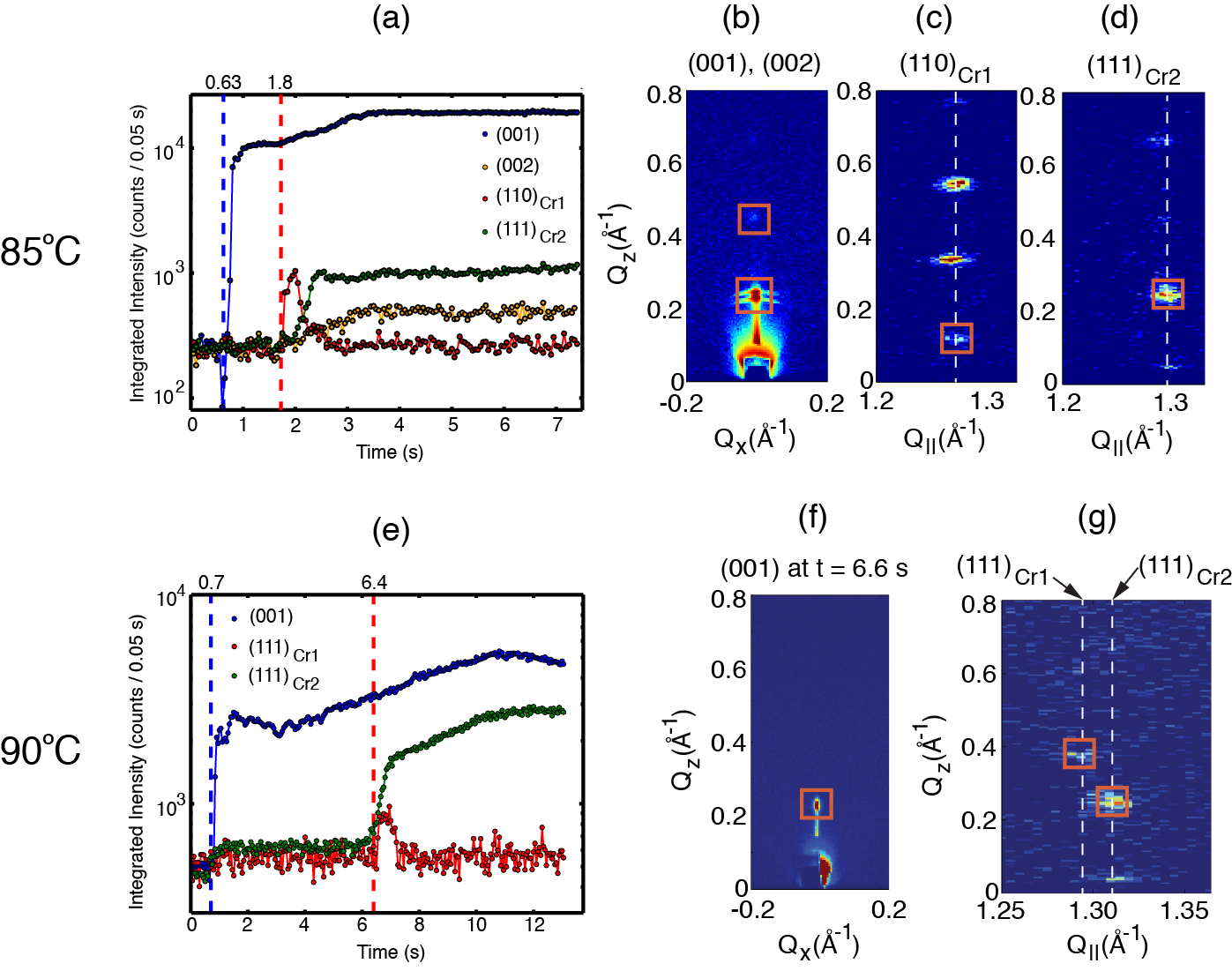}
\caption{\label{Intensity} Integrated Bragg peaks intensities during the film writing with a speed of 25 mm/s and substrate temperatures of (a) 85$^\circ$C and (c) 90$^\circ$C for a solution of 1wt$\%$ in toluene. The dashed lines indicate different stages of the crystallization process. At 0.63s in (a) and 0.7s in (e), the pen passes the X-ray beam, and thus the X-ray scattering pattern corresponds to the moment that the film is deposited. From $t=$0.63 s to 1.8 s in (a) and from 0.7 to 6.4 s in (e), (001) develops without any in-plane Bragg peaks, consistent with the LC phase. The second dashed in each case marks the beginning of the development of the in-plane structure.  (b)-(d) Êare GIWAXS images of the corresponding Bragg peaks intensities in (a). Boxes Êindicate the chosen areas for the integrated intensities. The $Q_{||}$ positions of the (11L) reflections are observed to shift from 1.28 {\AA}$^{-1}$ for Cr1 to 1.31 Ê{\AA}$^{-1}$ for Cr2. (g) shows a similar shift in $Q_{||}$ at 90$^\circ$ and also shows that (111) shifts in $Q_z$ from 0.38 Ê{\AA}$^{-1}$ to 0.25 {\AA}$^{-1}$. The image in (b) is at $t=$2.5 s. The times for (c) and (d) are 1.9 s and 2.4 s respectively, while (f) and (g) are at $t=$6.6 s. ÊThe complete sequences of images are shown in Supplementary Movies 2 (85$^\circ$C) and 3 (90$^\circ$C).\cite{Supplementary}}
Ê\end{figure*}

\section{Experiment}

C$_8$-BTBT was synthesized from commercially available 2-chlorobenzaldehyde via the methodology described by Takimiya et al.\cite{Ebata2007,Saito2011285}  Heavily doped n-type (100) silicon wafers with a 300 nm thermally grown silicon oxide layer were used as substrates for fabrication of bottom-gate, top-contact OFETs and for the X-ray measurements. We utilize the hollow capillary pen writing method with substrate heating to deposit thin films with controllable thickness and grain morphology.\cite{headrick2008} Real-time polarized optical microscopy (POM) was utilized to study the sequence of phases formed and the evolution of the grain morphology. In-situ microbeam grazing incidence wide-angle X-ray  scattering ($\mu$GIWAXS) was carried out for the same process at the Cornell High Energy Synchrotron Source (CHESS) at beamline D1. The X-ray wavelength was $\lambda$ = 1.155 {\AA}. A schematic of the layout is shown in Fig.~\ref{Expt_Layout}.\cite{Smilgies:2013aa} The X-ray intensities are plotted in Fig.~\ref{Expt_Layout}(b, c) as a function of the in-plane component of the wavevector transfer $Q_{||}$  and the component perpendicular to the substrate surface $Q_z$. Further details of the device fabrication and X-ray setup are given in the Supplementary Information file.\cite{Supplementary}

\section{Results and Discussion}

We have obtained a room temperature average OFET  mobility of  4.0 cm$^2$/V-s  for aligned C$_8$-BTBT thin films deposited in the convective regime at 0.5 mm/s (Supplementary Fig. S1).\cite{Supplementary}  This result is in good agreement with a previous report  of 3.5 - 5 cm$^2$/V-s by Uemura et al. for oriented films prepared by directed solution deposition on SiO$_2$/doped Si substrates.\cite{Uemura:2009aa} Remarkably, we find that nearly comparable results -- 2.7 cm$^2$/V-s  average and 3.0 cm$^2$/V-s  peak -- can be obtained in the LLD regime at 25 mm/s if the substrate is held at 60$^\circ$C. As we discuss below, real-time POM and $\mu$GIWAXS show that a series of transformations takes place that is consistent with Ostwald's rule of stages.\cite{ostwald1897studies} A key finding is that the grain size is controlled by the nucleation rate during the transformation from a transient liquid crystalline (LC) state to the crystalline form, and that the nucleation rate conforms to predictions of classical nucleation theory.\cite{porter_and_easterling}  For deposition temperatures up to 60$^\circ$C, LC transforms directly to the stable crystalline form Cr2 (see Supplementary Fig. S2).\cite{Supplementary} However, above 60$^\circ$C, a new intermediate crystalline state (Cr1) briefly forms, which leads to cracking of the films due to tensile strain as the final stable phase (Cr2) forms.

In Fig.~\ref{optical}, four frames from a polarized microscope movie for deposition at 90$^\circ$C substrate temperature are presented. Note that the substrate moves instead of the pen for these observations (while it is the pen that moves for the X-ray results). However,  it has already completed its motion by the time of Fig.~\ref{optical}(a) at  $\Delta t$ = 1.27 s relative to the time the area in view was written.  Fig.~\ref{optical}(a) shows that  a crystalline nucleus suddenly appears in the field of view. The drying of the film (not shown) occurs very fast, so the main part of the film visible in  Fig.~\ref{optical}(a), is interpreted as the LC phase.  The LC phase is only stable above 95$^\circ$C, so in this sequence it effectively exists in a supercooled state. A crystalline phase subsequently nucleates after a temperature-dependent incubation time. In Fig.~\ref{optical}(b) and (c) the crystalline grain expands in all directions.   Cracks are observed to progressively form in Fig.~\ref{optical}(c) and (d).  In Supplementary Movie 1,\cite{Supplementary}  the cracks are observed to sweep radially outward from the original nucleation center, which we interpret as the transition from Cr1 to Cr2.  As we discuss below, cracks degrade the performance of OFET devices even though the film has millimeter-scale grains. 

Fig.~\ref{Intensity}  shows the results of an in-situ crystallization study. The integrated intensities of the major Bragg peaks as a function of time during the deposition is shown in Fig. ~\ref{Intensity}(a) and (e) for films written at 25 mm/s at 85$^\circ$C and 90$^\circ$C respectively. The rapid growth of the  (001) reflection (blue data points and line in each case) starts almost right after the first dashed vertical line, which indicates $t$ = 0 when the X-ray beam starts to hit the film after the pen passes by it. Subsequently, the (001) reflection exists alone for quite some time.  During this time interval, the diffraction pattern is consistent with a smectic  liquid crystal  state, with layered structure perpendicular to the plane of the surface, but lacking any in-plane long-range order.\cite{gbabode2014}  Moreover,  we clearly observe that at 90$^\circ$C, the time interval of the LC state is longer than at 85$^\circ$C. The crystalline phases start to appear after the second vertical dashed line in Fig.~\ref{Intensity}(a) and (e). Note that  an intermediate crystalline phase Cr1 is briefly observed, but it transforms to Cr2 in less than one second.  Fig.~\ref{Intensity}(c, d, g) shows that Cr2 forms with the (11L) peaks shifted in both $Q_{||}$ and  $Q_z$ relative to Cr1. It is significant that Cr2 is at higher $Q_{||}$, which corresponds to a closer crystalline packing in the plane of the film.  This transition between solid phases evidently introduces a high ($\approx 2.4\%$) strain in the plane of the film that leads to mechanical failure and cracking of the films, as we have observed by POM in Fig.~\ref{optical}(d). The shift in $Q_z$ is interpreted as a shift in the stacking angle, coresponding to the $\beta$ angle of the unit cell.  After the film was cooled down to room temperature the positions of X-ray reflections were shifted due to thermal expansion (see Supplementary Fig. S3).\cite{Supplementary} The final positions are close to the bulk phase (see Supplementary Tables 1 $\&$ 2).\cite{Supplementary}

\begin{figure}
\includegraphics[width=2.0in]{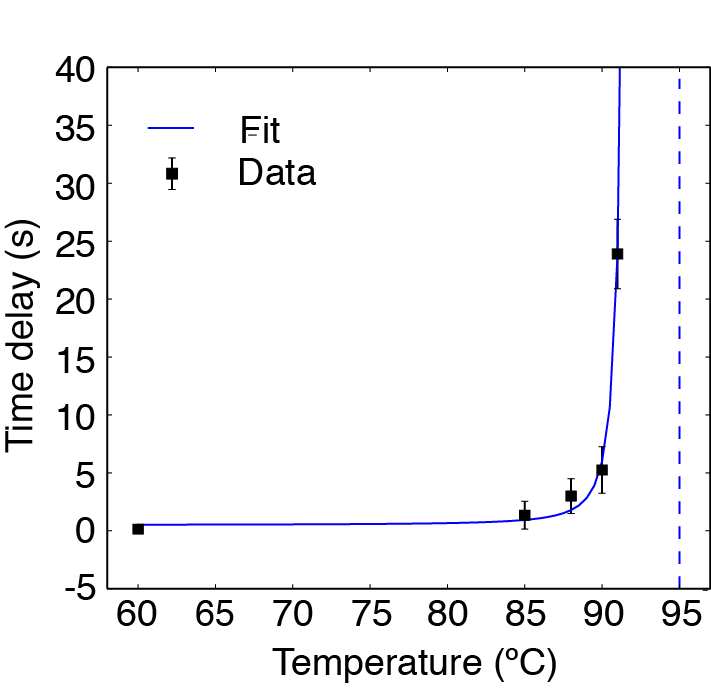}
\caption{\label{temp_delay}Time interval between the formation of the intermediate liquid crystal phase and the first appearance of (11L) in-plane at different substrate heating temperatures. When depositing at 95$^\circ$C, the film stays in the liquid crystal phase indefinitely. This is consistent with the transition temperature  for bulk samples observed by differential scanning calorimetry, as shown in Supplementary Fig. S4.\cite{Supplementary} }
\end{figure}

In Fig.~\ref{temp_delay}, we plot the time interval between the formation of the intermediate liquid crystal phase and the first appearance of (11L) in-plane reflections at different substrate heating temperature.  The trend is that it stays in the smectic phase longer at higher temperatures. We use a simple nucleation model to fit the data where the nucleation rate has the form $\exp(-\Delta G^*/k_BT)$, where $\Delta G^*$ is the Gibbs free energy nucleation barrier, $k_B$ is Boltzmann's constant, $T$ is the temperature.\cite{porter_and_easterling} Here, $\Delta G^*\propto 1/\Delta T^2$, where $\Delta T$ is the undercooling below the equilibrium transition temperature. This expression shows that when $\Delta T$ becomes small, the nucleation rate drops exponentially, and hence the incubation time increases rapidly.  But the crystallization velocity seems to be hardly affected, so that a low nucleation rate generally leads to a very large grain size.

Fig. 5 presents carrier mobility for deposition at 25 mm/s at different substrate temperature and the corresponding film morphology is shown in Supplementary Fig. S5.\cite{Supplementary}  The grain size increases with temperature, which correlates with the increasing mobility up to 60$^\circ$C where, as we have discussed above, no Cr1 phase is observed and thus there is no cracking related to the Cr1$\rightarrow$Cr2 phase transition. We have also found that  no cracking occurs for film thicknesses below 20 nm at 80$^\circ$C, which indicates that it is possible to stabilize the Cr1 metastable form (see Supplementary Fig. S6).\cite{Supplementary} 

The formation of transient phases during crystallization is an example of Ostwald's rule of stages,\cite{ostwald1897studies}  which is based on the empirical observation that  thermodynamically unstable phases often form before the stable  phase during crystallization from solution.  For such a sequence to occur, the transient phases must have lower activation barriers for nucleation compared to the stable phase,\cite{Chung:2009aa} where the $i$th nucleation barrier  $\Delta G^*_i$ originates from the interface energy at the boundaries between successive phases $i$ and $i+1$.   It is attractive to consider this transition in terms of reduced symmetry: for example, the LC phase is structurally similar to the isotropic  phase except for the loss of translational symmetry along the layering direction plus a degree of orientational ordering of the molecules. Similarly, the crystalline phase loses translational symmetry in the plane of the layers, and thus becomes even less similar to the original isotropic state. Thus, it is reasonable that the isotropic$\rightarrow$LC transition has a lower nucleation barrier  compared to the direct  isotropic$\rightarrow$Cr transition because more similar structures should have comparatively lower interface energies between them. Based on this argument, the  configurational entropy of the system is also lowered, so the order of appearance of phases more fundamentally follows a rule of decreasing entropy, where the phase formed at each stage is the one with the smallest entropy change.  One prediction of this rule is that any material that has an LC phase should transform to the LC phase before reaching its crystalline form. \\

\begin{figure}
\includegraphics[width=3.0 in]{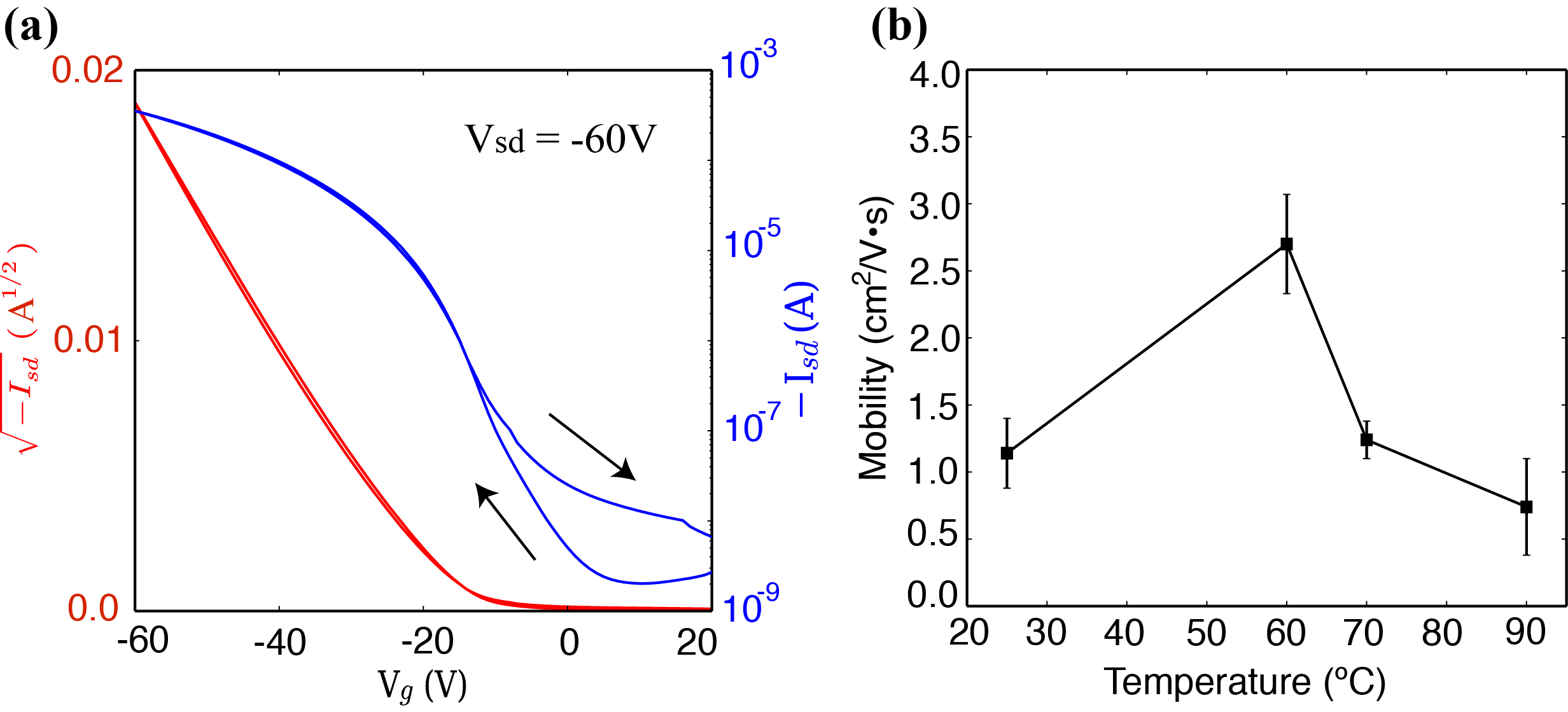}
\caption{\label{temp_mobility}(a) Drain current $I_d$ versus gate voltage $V_g$ transfer curve for one of the transistors fabricated at 60$^\circ$C and 25 mm/s. The characteristics show $I_d$ on a logarithmic scale on the right (blue with direction arrows) and $\sqrt{-I_d}$ on a linear scale on the left (red) versus $V_g$. The mobility extracted from this data is 3.04 cm$^2$/V-s. (b) OFET average saturation mobility versus substrate heating temperature for a C$_8$-BTBT solution of 0.6 wt$\%$ concentration at a writing speed 25 mm/s.} 
\end{figure}



%

\section{Conclusions}
We have described the formation of transient phases of C$_8$-BTBT thin films during solution processing at high speed.  Although striking and somewhat surprising, the results of POM, in-situ $\mu$GIWAXS, and OFET studies point to a model that is entirely consistent with  classical nucleation theory and with Ostwald's rule.  These results lead to a method to produce extremely large grain size films and high carrier mobility in the LLD regime that may prove to be of considerable practical importance. We predict that many additional materials will be found to exhibit high-symmetry transient phases, particularly those that can form an LC phase.\\

\begin{acknowledgments}
This work was supported by the National Science Foundation, Division of Materials Research,  Electronic and Photonic Materials Program through award DMR-1307017. The X-ray scattering research was conducted at the Cornell High Energy Synchrotron Source (CHESS) which is supported by the NSF and the National Institutes of Health/National Institute of General Medical Sciences under NSF award DMR-1332208.
\end{acknowledgments}

\bibliography{Transient_phases_V3b}

\end{document}